\documentclass [aps,pra,amssymb,amsmath,twocolumn, showpacs]{revtex4-1}
\usepackage[latin9]{inputenc}
\usepackage{amsmath}
\usepackage{amssymb}
\usepackage{graphicx}
\usepackage{verbatim}
\usepackage{bm}
\usepackage{color}

\usepackage{graphicx,epsfig,amsfonts,amssymb}
\usepackage{bm}
\usepackage{times}
\usepackage{lipsum}
\usepackage{verbatim}

\newcommand{\nn}{\nonumber}

\newcommand{\la}{\langle}
\newcommand{\ra}{\rangle}
\newcommand{\rar}{\rightarrow}
\newcommand{\da}{\downarrow}
\newcommand{\ua}{\uparrow}
\newcommand{\be}{\begin{eqnarray}}
\newcommand{\ee}{\end{eqnarray}}

\newcommand{\bs}{\begin{equation}\begin{split}}
\newcommand{\es}{\end{split}\end{equation}}

\begin{document}
\title{Landau-Zener extension of the Tavis-Cummings model: structure of the solution}
\author{Chen Sun$^{a,b}$}
\email{chen.sun.whu@gmail.com}


\author{Nikolai A. Sinitsyn$^{b}$}
\email{nsinitsyn@lanl.gov}
\affiliation{$^a$Department of Physics, Texas A\&M University, College Station, TX 77843, USA}
\affiliation{$^b$Theoretical Division, Los Alamos National Laboratory, Los Alamos, NM 87545, USA}

\begin{abstract}
We explore the recently discovered solution of the driven Tavis-Cummings model (DTCM). It describes interaction of arbitrary number of two-level systems with a bosonic mode that has linearly time-dependent frequency. We derive compact and tractable expressions for transition probabilities in terms of the well known special functions. In the new form,  our formulas are suitable for  fast numerical  calculations and analytical approximations. As an application, we obtain the semiclassical limit of the exact solution and compare it to prior approximations.  We also reveal  connection between DTCM and  $q$-deformed binomial statistics.
\end{abstract}

\maketitle



\section{Introduction}


Throughout its history, quantum physics has been strongly influenced by the discovery of
 exact results, i.e. analytical expressions derived without approximations and consequently valid at arbitrary values of model parameters.
 In stationary quantum mechanics, there is a wide range of highly useful exactly solvable models, from quantum oscillator and  hydrogen atom to models with truly many-body interactions, such as the 1D Hubbard model.

An important goal in physics is to achieve control over quantum  dynamics.
This  can be done by application of explicitly time-dependent fields to the considered quantum system.
Unfortunately, unlike stationary quantum mechanics, the class of useful exact results for  quantum models with time-dependent parameters is very restricted.
 Most known models of this class are elementary, e.g.,  the quantum harmonic oscillator with time-dependent force  or   two-state systems with specially chosen time-dependent parameters \cite{Berry_2009}. This scarcity of nontrivial exact results   restricts our ability to understand and control quantum dynamics. Hence, more advanced exact solutions in nonstationary quantum mechanics are  needed.

It has been shown \cite{mlz-all} that considerable progress can  be achieved in solving, so-called, multistate Landau-Zener (LZ) models \cite{majorana}, which Hamiltonians have the form
\begin{equation}
 \hat{H}(t) = \hat{A} +\hat{B}t .
\label{mlz}
\end{equation}
Here, $\hat{A}$ and $\hat{B}$ are constant Hermitian $N\times N$ matrices.  One can always choose the, so-called, {\it diabatic basis} in which the matrix $\hat{B}$ is diagonal,
and if any pair of its elements are degenerate then the corresponding off-diagonal element of the matrix $\hat{A}$ can be set to zero by a time-independent change of the basis, that is
\be
B_{ij}= \delta_{ij}\beta_i, \quad  A_{nm}=0\,\,\, {\rm if} \,\, \beta_{n}=\beta_{m}. 
\label{diab3}
\ee
Constant parameters $\beta_i$  are called the {\it slopes of diabatic levels},  diagonal elements of the Hamiltonian in the diabatic basis,
$B_{ii}t+A_{ii}$, are called {\it diabatic energies}, and  nonzero off-diagonal elements of the matrix $\hat{A}$ in the diabatic basis are called the {\it coupling constants}.
Usually, evolution at arbitrary time  is impossible to obtain analytically. Instead, the goal of the multistate LZ theory is to find the scattering $N\times N$ matrix $\hat{S}$, whose element $S_{nn'}$ is the amplitude of the diabatic state $n$ at $t  \rightarrow +\infty$, given that at $t \rightarrow -\infty$ the system was in the $n'$-th diabatic state. In most cases, only the related matrix $\hat{P}$, with elements $P_{n' \rightarrow n}\equiv |S_{nn'}|^2$,  called the {\it matrix of transition probabilities}, is needed.


Recently, a truly many-body model of the type (\ref{mlz}) was solved exactly in terms of an algorithm  that leads, in a finite number of steps, to exact expression for any element of the transition probability matrix.
The  model solved in \cite{Sinitsyn_2016} corresponds to the time-dependent version of the generalized Tavis-Cummings system with $N_s$ spins and the Hamiltonian
\begin{align}
\label{HamiltonianTC}
\hat{H}(t)=t\hat{a}^\dag \hat{a}+\sum_{i=1}^{N_s}\epsilon_i\hat{\sigma}_i+g\sum_{i=1}^{N_s}(\hat{a}^\dag \hat{\sigma}_i^-+\hat{a} \hat{\sigma}_i^+),
\end{align}
where $t$ is time, $\hat{a}$ is the boson annihilation operator, $\hat{\sigma}_i^{\pm}$ are the $i$th spin's raising and lowering operators, $g$ describes the coupling of spins to  bosons, $\epsilon_i$ is the intrinsic level splitting of the $i$th spin, and
$
\hat{\sigma}_i\equiv(\hat{1}+\hat{\sigma}_z^i)/2
$
 is the projection operator to spin ``up'' state of the $i$th spin, where $\hat{1}_i$ is a unit matrix acting in the $i$th spin subspace, and $\sigma_z$ is the Pauli $z$-matrix of the $i$th spin.

The degenerate case of Eq.~(\ref{HamiltonianTC}) with $\epsilon_i=0$ for all $i=1,2,\ldots,N_s$ is also of interest. After decoupling all non-symmetric states, it is equivalent to the following time-dependent  model with the Hamiltonian
\begin{align}\label{Hamiltonian}
\hat{H}(t)=t\hat{a}^\dag \hat{a}+g\left(\hat{a}^\dag \hat{S}^-+\hat{a} \hat{S}^+ \right),
\end{align}
where $\hat{S}^{\pm}$ are the spin raising/lowering operators with a spin size $S=N_s/2$.

Models (\ref{HamiltonianTC}) and (\ref{Hamiltonian}) with linear time-dependence of the bosonic mode frequency describe  an  important process of conversion of ultracold fermionic atoms into a molecular condensate during a linear sweep of the magnetic field through the Feshbach resonance \cite{yurovski}. Hence, even before the finding of the exact solution, different approximations had been developed to understand models  (\ref{HamiltonianTC}) and (\ref{Hamiltonian}), including perturbative calculations for small coupling \cite{dobrescu}, diagrammatic kinetic approach \cite{Altland_2008,Altland_2009}, semiclassical approximations \cite{Altland_2009,Itin_2009,Itin_2010}, and
simplifying the models in particular limits by mapping them to the known solvable cases of the type (\ref{mlz}) \cite{limits-LZ}. Discovery of the most general exact solution provides the possibility to explore regimes when all such approximations are not applicable. It is also tempting to verify validity of the previously developed approximate methods by comparing them to exact formulas.

However, in Ref.~\cite{Sinitsyn_2016}, solution of the models (\ref{HamiltonianTC}) and (\ref{Hamiltonian}) was presented in the form of a procedure that required examination of  semiclassical trajectories, which number is growing exponentially with $N_s$. So, even when it was possible to write compact expressions for some matrix elements, useful results were expressed through multiple sums over quickly growing number of terms. Therefore, solution in \cite{Sinitsyn_2016} was not very suitable for direct numerical calculations or comparison to previously known estimates.

In this article, we explore  exact solution \cite{Sinitsyn_2016} with the goal to transform it to a more tractable form. Achieving this goal allows us to study behavior of  transition probabilities at large numbers of spins, and look at various other limits  for direct comparison with prior theoretical predictions.  In section 2, we derive a simple formula for arbitrary element of the transition probability matrix for the model  (\ref{HamiltonianTC}). In section 3, we consider physically most important case when all spins (or the arbitrary spin $S$ in the case of model (\ref{Hamiltonian})) are, initially, fully polarized and derive expressions for  probabilities to change the total polarization in terms of $q$-Pochhammer symbols. In section 4, we  consider the limit of a large number of spins and derive continuous approximation, which we compare with previously derived approximate solutions of the models  (\ref{HamiltonianTC}) and (\ref{Hamiltonian}) in the semiclassical limit. We will summarize our findings in the conclusion.


\section{State-to-state transition probabilities}
 In \cite{Sinitsyn_2016}, solution of the model \eqref{HamiltonianTC} was presented in terms of   semiclassical paths in some time-energy diagram. Such a form of a solution is common for all other solvable multistate LZ models, however, we are going to show  that it is not optimal for DTCM. Our new derivation  leads to a simple analytical formula for probabilities of all possible elements of the transition probability matrix of a model with an arbitrary number of spins.

First, we recall that the number of excitations, i.e., the number of bosons plus the number of spins up is conserved:
\begin{equation}
N_e\equiv \hat{a}^{\dagger} \hat{a} +\sum_{i} \hat\sigma_i = {\rm const}.
\label{nex}
\end{equation}
Hence, the Hamiltonian \eqref{HamiltonianTC} can be rewritten as
\begin{align}
\label{HTC}
\hat{H}(t)=\sum_{i=1}^{N_s}(\epsilon_i-t) \hat{\sigma}_i+g\sum_{i=1}^{N_s}(\hat{a}^\dag \hat{\sigma}_i^-+\hat{a} \hat{\sigma}_i^+),
\end{align}
where we disregarded the constant that depends on $N_e$ as it does not influence the dynamics.
We represent $2^{N_s}$ diabatic states of  model (\ref{HTC}) by a vector of zeros and ones $|\sigma_1,\sigma_2,\ldots,\sigma_{N_s}\ra$, where $\sigma_i$ being $1$ or $0$ corresponds to $i$th spin being $\ua$ or $\da$ along $z$-axis. Diabatic energy of such a state is given by $\sum_{i=1}^{N_s}(\epsilon_i-t) \sigma_i$. We will assume that $\epsilon_1>\epsilon_2>\ldots>\epsilon_{N_s}$.

According to \cite{Sinitsyn_2016}, exact solution for transition probabilities between diabatic states in the model (\ref{HTC}) coincides with the result of a simple stochastic process in which transitions between diabatic levels happen only at moments of level intersections. Pairwise transition probabilities are then determined by a simple LZ formula. As time goes from $-\infty$ to $\infty$, there will be moments  at
$
t_i=\epsilon_i
$,
at which two diabatic states with $i$-th spin projection ``up" and ``down" have the same diabatic energy. This means that, chronologically, the spin with the lowest splitting $\epsilon_{N_s}$ will be first to encounter the moment when it can flip, the spin with energy $\epsilon_{N_s-1}$ will be the second, and so on. There will be, totally, exactly $N_s$ such moments.

Pairwise transition probabilities will depend on the number of bosons at a given intersection.
If $N_B$ is the number of bosons in the state with all spins polarized ``up'', then the states $|\sigma_1,\sigma_2,\ldots,\sigma_j=0,\ldots,\sigma_{N_s}\ra$ and $|\sigma_1,\sigma_2,\ldots,\sigma_j=1,\ldots,\sigma_{N_s}\ra$ will have $N_B+ k $ and $N_B+k-1$ bosons, respectively, where $k=N_s-\sum_{i=1,i\ne j}^{N_s}\sigma_i$. Corresponding coupling between these two diabatic states is $g_{k}=g\sqrt{N_B+k}$, the probability to stay on the same level is
$p_{k}=e^{-2\pi g^2(N_B+k)}$, where $k=1,\ldots,N_s$;  and the probability to turn to the other diabatic state is $q_{k}=1-p_{k}$.
We chose indexes of $p$ and $q$ to start with 1 in order to be consistent with notation in Ref.~\cite{Sinitsyn_2016}.
 Thus, we can summarize the process of deriving transition probability between any pair of diabatic states in the form of the following algorithm:


1. A transition from an initial state $|I\ra=|\sigma^I_1,\sigma^I_2,\ldots,\sigma^I_{N_s}\ra$ to a final state $|F\ra=|\sigma^F_1,\sigma^F_2,\ldots,\sigma^F_{N_s}\ra$ consists of $N_s$ steps, each step being a flip or stay of a single spin. The order of spin flips/stays should be from the right to the left, i.e., the first step corresponds to the spin $\sigma_{N_s}$, the second step corresponds to $\sigma_{N_s-1}$, etc. Each flip or stay generates some factor $q_{k}$ or $p_k$, respectively.

2. The subscript $k$ of a factor $q_k$ or $p_k$ is determined by the \textit{transient} spin configuration at its corresponding step -- it equals 1 plus the number of down ($\da$) spins for all the $N_s$ spins \textit{except} the spin involved in that step.

3. Process terminates after finding a corresponding factor, $q_k$ or $p_k$, for the first spin. The final transition probability $P_{I\rar F}$ is the product of all such factors from all spins.

As an example, consider the transition from the state $|I\ra=|\ua,\da,\ua\ra$ to the state $|F\ra=|\da,\da,\da\ra$ in $N_s=3$ case. This process should be decomposed into three steps, each of them being a flip/stay of a single spin, from the rightmost spin to left. The process is illustrated in the following table, where the spin involved in each step is marked as a double arrow:


\bigskip

\begin{tabular}{c|c|c|c|c}
\hline
step & spin involved  & flip/stay & change of states & factor\\
\hline
1 & $\sigma_3$ & flip & $|\ua,\da,\Uparrow\ra \rar |\ua,\da,\Downarrow\ra$ & $q_2$\\
2 & $\sigma_2$ & stay & $|\ua,\Downarrow,\da\ra \rar |\ua,\Downarrow,\da\ra$ & $p_2$ \\
3 & $\sigma_1$ & flip & $|\Uparrow,\da,\da\ra\rar |\Downarrow,\da,\da\ra$ & $q_3$ \\
\hline
\end{tabular}

\bigskip

\noindent The  probability of this transition can then be read out as $P_{I\rar F}=q_2p_2q_3$.

From the above procedure, we can write an expression for transition probabilities from an initial state
$|I\ra=|\sigma^I_1,\sigma^I_2,\ldots,\sigma^I_{N_s}\ra$ to a final state $|F\ra=|\sigma^F_1,\sigma^F_2,\ldots,\sigma^F_{N_s}\ra$ (recall that $\ua=1$, $\da=0$):
\begin{align}\label{2}
P_{I\rar F}=\prod_{i=1}^{N_s} p^{\delta_{\sigma^I_i,\sigma^F_i}}_{k_i}q^{1-\delta_{\sigma^I_i,\sigma^F_i}}_{k_i},
\end{align}
where the subscript reads:
\begin{align}\label{3}
k_i={N_s-\sum_{j=1}^{i-1}\sigma^I_j-\sum_{j=i+1}^{N_s}\sigma^F_j}.
\end{align}
The exponents containing Kronecker deltas serve to determine whether $p$ or $q$ is included in a single step: if $\sigma^I_i=\sigma^F_i$, $p$ is included, and if $\sigma^I_i\ne\sigma^F_i$, $q$ is included.
\begin{figure}[!htb]
\scalebox{0.325}[0.325]{\includegraphics{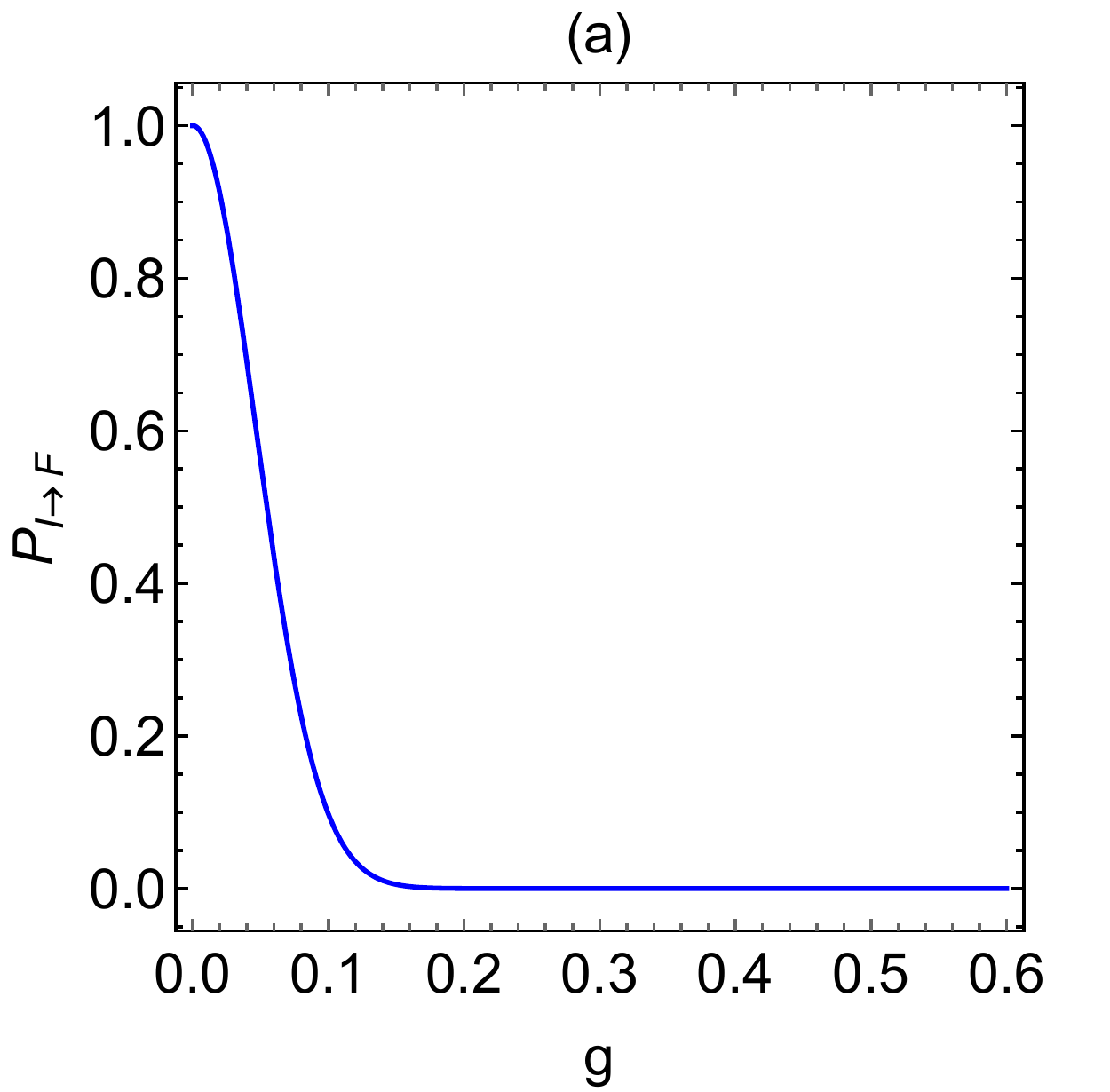}}
\scalebox{0.345}[0.345]{\includegraphics{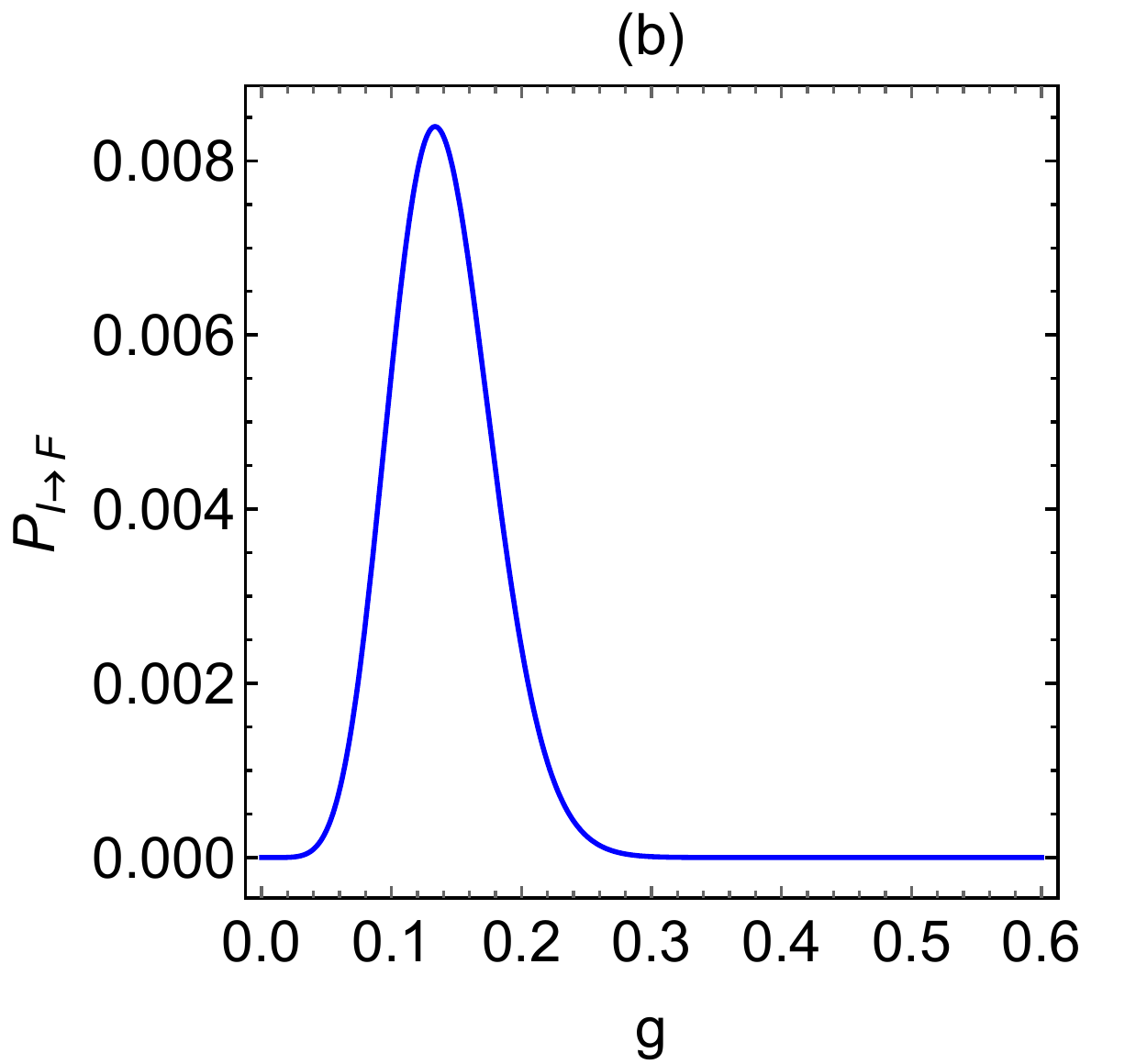}}
\scalebox{0.345}[0.345]{\includegraphics{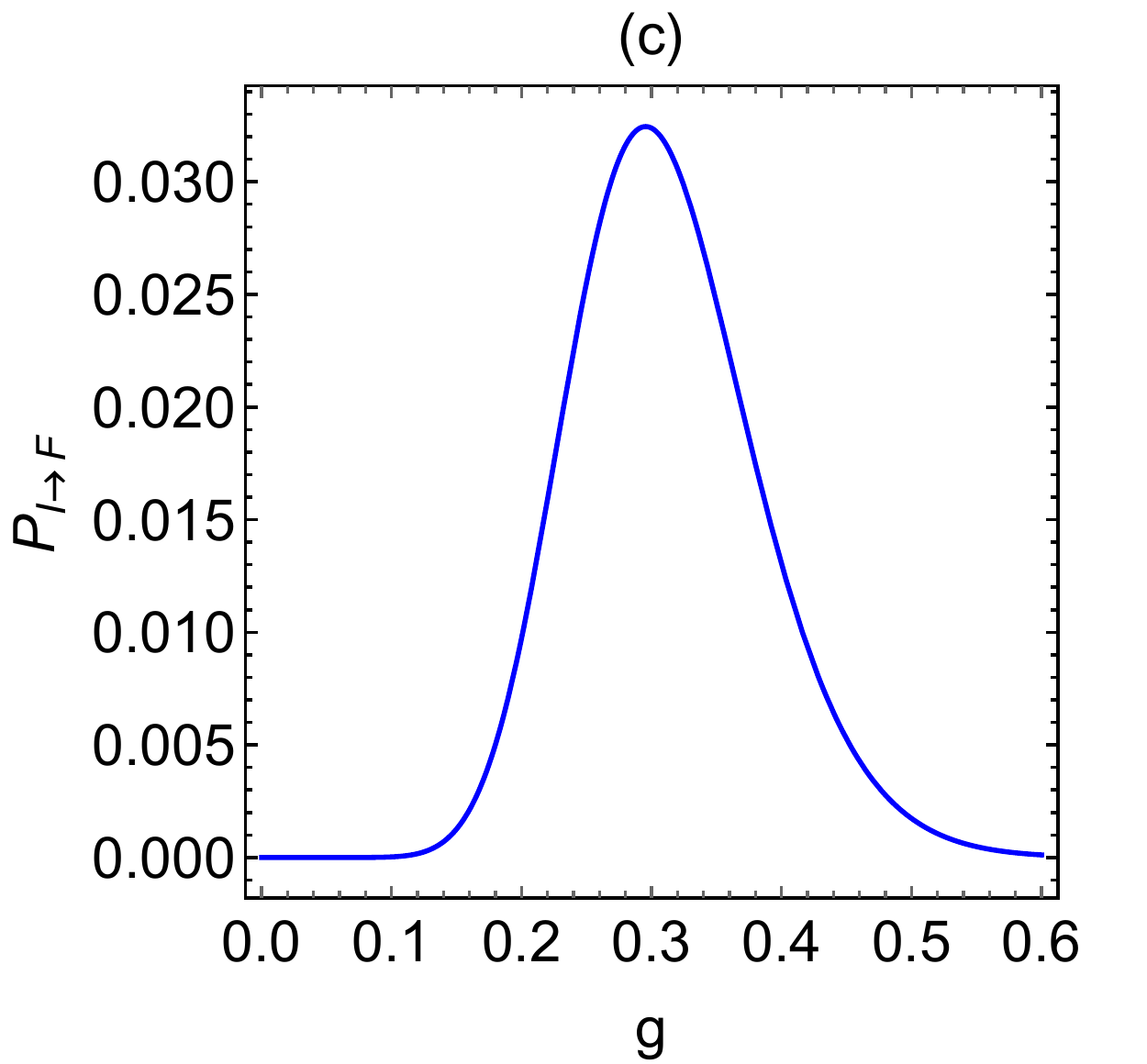}}
\scalebox{0.325}[0.325]{\includegraphics{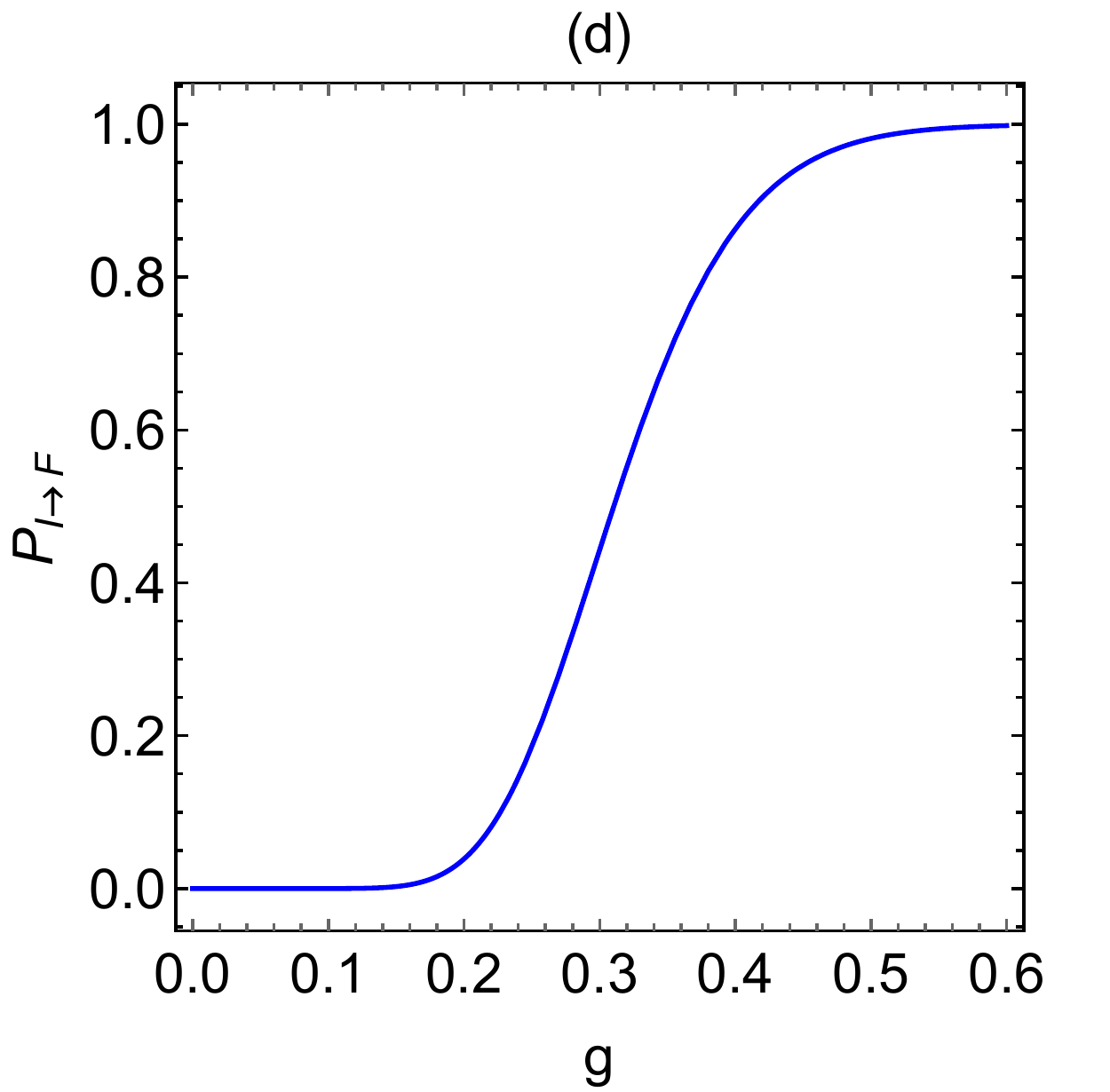}}
\caption{Samples of transition probabilities $P_{I\rar F}$ vs. $g$ predicted by Eq.~\eqref{2} for $N_s=7$ at $N_B=0$. In all cases, $I=|\da,\da,\ua,\da,\ua,\da,\da\ra$ and  (a) $F=|\da,\da,\ua,\da,\ua,\da,\da\ra$ (the same state); (b) $F=|\da,\ua,\ua,\da,\da,\ua,\da\ra$; (c) $F=|\ua,\ua,\da,\ua,\ua,\ua,\ua\ra$; (d) $F=|\ua,\ua,\da,\ua,\da,\ua,\ua\ra$ (the ``opposite spin'' state). For these transitions, Eq.~\eqref{2} provides the following probabilities: (a) $p_5^5 p_6^2$; (b) $p_4p_5^2p_6q_5^3$; (c) $p_4 q_2 q_3^3 q_4 q_5$; (d) $ q_3 q_4^5 q_5$. Among them, (a) represents the main diagonal elements of the transition probability matrix, and (d) represents the anti-diagonal elements.
}
\label{pulses}
\end{figure}

It is not difficult to prove that our Eq.~\eqref{2} satisfies the symmetry properties of the solution described in \cite{Sinitsyn_2016}.
For example, the property (28) in \cite{Sinitsyn_2016} describes a symmetry between the transitions $|I\ra=|\sigma^I_1,\sigma^I_2,\ldots,\sigma^I_{N_s}\ra\rar|F\ra=|\sigma^F_1,\sigma^F_2,\ldots,\sigma^F_{N_s}\ra$ and $|\bar I\ra=|\bar\sigma^I_1,\bar\sigma^I_2,\ldots,\bar\sigma^I_{N_s}\ra\rar|\bar F\ra =|\bar\sigma^F_1,\bar\sigma^F_2,\ldots,\bar\sigma^F_{N_s}\ra$, where $\bar\sigma_i\equiv1-\sigma_i$. 
In Eq.~\eqref{2} we see that all the exponents are the same for these two transitions, since all the delta functions remain unchanged. But a subscript $k_i$ for the first transition will become $N_s-\sum_{j=1}^{i-1}(1-\sigma^I_j)-\sum_{j=i+1}^{N_s}(1-\sigma^F_j)=N_s+1-k_i$ for the second transition. Equation~\eqref{2} then predicts that, under exchanges of $p_k$ with $p_{N_s+1-k}$ and $q_k$ with $q_{N_s+1-k}$ for all $k=1,\ldots,N_s$, one transition probability changes to the other. Thus, Eq.~\eqref{2} satisfies the property 
$P_{I \rar F}(p_1,\ldots,p_{N_s},q_1,\ldots, q_{N_s})=P_{\bar I \rar \bar F}(p_{N_s},\ldots,p_{1},q_{N_s},\ldots, q_{1})$. Equation~(\ref{2}) also reproduces correctly all formulas that were derived in \cite{Sinitsyn_2016}, e.g., the explicit solution for the three spin sector of the model (\ref{Hamiltonian}).

Finally, we use Eq.~\eqref{2} to get a look at typical behavior of pairwise transition probabilities for a model with $N_s=7$ spins. For such a large number of spins, direct numerical solution of the nonstationary Schr\"odinger equation is very difficult to achieve with reasonable precision, however, Eq.~\eqref{2} provides a quick answer.  Figures~\ref{pulses}(a) and (d) show transition probability matrix elements along, respectively, first and second  main diagonals. Such elements depend on products of only, respectively, $p$ and $q$ factors, which explains their monotonic dependence on coupling $g$. Figures~\ref{pulses}(b) and (c) show typical behavior of off-diagonal elements of the transition probability matrix. Such elements are given by a product of  some number of monotonically decreasing $p$-factors times some number of  monotonically growing  $q$-factors. Not surprisingly, the resulting transition probability has the shape of a pulse.

\section{Probability distribution of total change of spin polarization}

The number of possible diabatic states is growing exponentially with the number of spins. So, realistic applications of the model (\ref{HamiltonianTC})  require the knowledge of coarse-grained characteristics, such as the number of emitted bosons or, equivalently, the change of the total spin polarization regardless  which spins where flipped. Also, the  most interesting for practical reasons initial conditions  are such that $N_s\equiv 2S$  spins in the model (\ref{HamiltonianTC}) are initially either all polarized ``up" or all polarized ``down" with some initial number of bosons being, respectively, $N_B$ or $N_B+2S$. It was shown in \cite{Sinitsyn_2016} for  model (\ref{HamiltonianTC}) that the sum of transition probabilities from one  of such  fully polarized states  to all possible  states with $\nu$ spins pointing down  is given by one of the corresponding formulas:





\begin{align}
&P_{0\rar \nu}=\left(\prod_{k=1}^{\nu}(1-p_k) \right)\sum_{i_1,\ldots,i_{\nu+1}=0}^{2S-\nu}\delta_{\sum_{k=1}^{\nu+1}i_k,2S-\nu}\prod_{r=1}^{\nu+1}p_r^{i_r},\label{exactP1n}\\
&P_{2S\rar \nu}=\left( \prod_{k=\nu+1}^{2S}(1-p_k)  \right)\sum_{i_{\nu},\ldots,i_{2S}=0}^{\nu}\delta_{\sum_{k={\nu}}^{2S}i_k,\nu}\prod_{r=\nu}^{2S}p_r^{i_r},\label{exactP2Sp1n}
\end{align}
where
\begin{equation}
p_k=e^{-2\pi g^2(N_B+k)}, \quad k\in (1,\ldots,2S).
\label{pk}
\end{equation}
Here, the state with all spins ``up" has index $0$ and the state with all spins ``down" has index $2S$. We write $2S$ instead of $N_s$ because it was shown in \cite{Sinitsyn_2016}
that formulas (\ref{exactP1n})-(\ref{exactP2Sp1n}) also describe state-to-state transition probabilities in the degenerate model (\ref{Hamiltonian}) with $2S+1$ states, where index $\nu$ corresponds to spin projection $S_z=S-\nu$. Later in this article, we will compare our results with results derived specifically for  model (\ref{Hamiltonian}) with similar definition of state indexes as in (\ref{exactP1n})-(\ref{exactP2Sp1n}), so it is more convenient to use terminology of the degenerate model.

We will refer to formulas (\ref{exactP1n}) and (\ref{exactP2Sp1n}) as describing, respectively, forward and inverse  processes.
The Hamiltonians  (\ref{HamiltonianTC})-(\ref{Hamiltonian}) conserve the number of excitations, i.e., the number of spins ``up" plus the number of bosons. Therefore, one can interpret   $P_{0\rar \nu}$ as the probability to emit $\nu$ bosons when starting with state $\nu'=0$, and  $P_{2S\rar \nu}$ as the probability to absorb $2S-\nu$ bosons when starting with state $\nu' = 2S$.

Let us now introduce the $q$-Pochhammer symbol:
\begin{align}\label{qPoch}
(a,q)_k=\prod_{i=0}^{k-1}(1-a q^i).
\end{align}
In Eqs.~\eqref{exactP1n}-\eqref{exactP2Sp1n} the products of $(1-p_k)$'s, with  definition (\ref{pk}), can be directly identified as $q$-Pochhammer symbols.
Next we note that the sums in Eqs.~\eqref{exactP1n}-\eqref{exactP2Sp1n} are similar to the canonical partition function of some $N$ free bosons populating $n$ equidistant  quantum energy levels:
\begin{align}\label{partition}
Z_N=\sum_{i_1,i_2,\ldots,i_n=0}^{N}\delta_{i_1+i_2+\ldots+i_n,N}\prod_{r=1}^{n}e^{-\beta (r-1)  i_r},
\end{align}
where $\beta=1/k_B T$ is the inverse temperature and   $i_r$ is the number of bosons on the level with index $r$. Here we assumed that
the energy difference $\Delta$ between  the nearest levels is normalized to $\Delta=1$.

We will calculate the partition function in (\ref{partition}) following the method described in \cite{Mullin_2003}:
First, consider the well known grand canonical partition function of free bosons that can occupy $n$ equidistant energy levels:
$\mathcal{Z}(z)=\sum^\infty_{N=0}Z_Nz^N$, where $z=e^{\beta \mu}$, with $\mu$ being chemical potential:
\begin{align}\label{calZ}
\mathcal{Z}(z)=\prod_{k=1}^n\frac{1}{1-zx^{k-1}},
\end{align}
where $x=e^{-\beta }$. On the other hand, we can formally expand
 \begin{equation}
 Z_N=\sum_{M=0}^{N(n-1)}C_N(M)x^M,
\label{zexp1}
\end{equation}
where $M$ labels integer valued energies of the system and $C_N(M)$ is the number  of degenerate states
with total energy  $M$. The grand canonical partition function can then be written in the form
\begin{align}\label{calZexpand}
\mathcal{Z}(z)=\sum_{N}\sum_{M}C_N(M)x^Mz^N.
\end{align}
Here, we observe that if we
replace $z$ in $\mathcal{Z}(z)$, in Eq.~(\ref{calZ}), with $xz$, we find
\begin{align}\label{recursion}
(1-zx^{n})\mathcal{Z}(xz)=(1-z)\mathcal{Z}(z).
\end{align}
Combining this with  Eq.~\eqref{calZexpand} leads to
\begin{align}\label{}
&\sum_{N}\sum_{M}C_N(M)x^{M+N}z^N-\sum_{N}\sum_{M}C_N(M)x^{M+N+n}z^{N+1}\nn\\
&=\sum_{N}\sum_{M}C_N(M)x^{M}z^{N}-\sum_{N}\sum_{M}C_N(M)x^{M}z^{N+1}.
\end{align}
Equating the same powers of $z$ and
using 
(\ref{zexp1}), we get
\begin{align}\label{}
&x^{N}Z_N-x^{N+n-1}Z_{N-1}=Z_N-Z_{N-1},
\end{align}
which leads to the recursion relation for $Z_N$:
\begin{align}\label{}
&Z_N=\frac{1-x^{N+n-1}}{1-x^N}Z_{N-1}.
\end{align}
With the starting value $Z_0=1$, this leads to a closed expression \cite{note-1}:
\begin{align}
\label{note-Z}
&Z_N=\prod_{k=1}^N\frac{1-x^{k+n-1}}{1-x^k}=\frac{(x^{n};x)_N}{(x;x)_N}.
\end{align}
 Finally, we note that if  level energies start not from 0 but from some value $E$, and  increase in unit steps, then
such a uniform shift of all level energies  merely introduces an overall additional factor $e^{-\beta N E}$ to the partition function.


Comparing the sums in Eqs.~\eqref{exactP1n}-\eqref{exactP2Sp1n} with Eq.~(\ref{partition}) we find that sums in  $P_{0\rar\nu}$ correspond to a system with $2S-\nu$ bosons on $\nu+1$ levels, with energy of  the lowest level equals $N_B+1$, while the sums in $P_{2S\rar\nu}$ correspond to a system with  $\nu$ bosons on $2S+1-\nu$ levels, and energy of the lowest level equals $N_B+\nu$.
In both cases we should identify $\beta=2\pi g^2$.

Let us denote
$$
x\equiv e^{-2\pi g^2},
$$
 and introduce the $q$-binomial coefficients:
 \begin{align}
\left[\begin{array}{c}n\\k\end{array}\right]_x=\frac{(x;x)_n}{(x;x)_k(x;x)_{n-k}},
\end{align}
in terms of which we can now express transition probabilities from fully polarized states:
\begin{align}
&P_{0\rar \nu}
=\left[\begin{array}{c}
2S\\\nu
\end{array}\right]_x x^{(N_B+1)(2S-\nu)}( x^{N_B+1};x)_\nu,\label{q0nu}\\
&P_{2S\rar \nu}
=\left[\begin{array}{c}
2S\\\nu
\end{array}\right]_xx^{(N_B+\nu)\nu}(x^{N_B+\nu+1};x)_{2S-\nu},\label{q2Snu}
\end{align}

One immediate utility of rewriting solutions in the form (\ref{q0nu})-(\ref{q2Snu}) is that numerical time of Pochhammer symbol calculation scales linearly with $S$, so we can easily find numerically exact values of transition probabilities for, e.g., $S\sim10^3$.
Second, Pochhammer symbol is a well known special function. Its properties, however complex, have been extensively investigated, including asymptotic behavior at some limits of  parameters.  Finally, it is known that Pochhammer symbol usually emerges in physical applications in relation to quantum algebras  \cite{q-algebra} which operate with deformations of physical characteristics, including statistical distributions.  Such deformations are usually described by an operator algebra that depends on a continuous parameter $q$, such that at $q=1$ standard, e.g. a Lie group, physical relations among operators are recovered.

 The latter property hints on the origin of the exact solution, which is currently lacking mathematically rigorous justification. An additional indication to that solution of DTCM can be related to a $q$-deformed quantum algebra follows from the observation that the distribution in Eq.~(\ref{q0nu}) is actually
the $q$-deformed binomial distribution \cite{Jing_1994,Chung_1995}, which is also known as the $q$-Bernstein basis
 \cite{Phillips_1997,Phillips_2010}. This distribution is formally defined as
\begin{align}
B_k^n(\tau;q)=\left[\begin{array}{c}
n\\k\end{array}\right]_q \tau^k (\tau;q)_{n-k}.
\label{bbk}
\end{align}
with some parameters $\tau$ and $q$.
Hence, we can also write
\begin{align}\label{}
P_{0\rar \nu}=B_{2S-\nu}^{2S}(x^{N_B+1};x).
\end{align}
Since its introduction \cite{Jing_1994}, this distribution has been extensively studied. For example, its generating function is given in \cite{Jing_1994}, and its mean can be expressed in terms of the $q$-Pochhammer symbols and $q$-binomial coefficients \cite{Zeiner_2010}. Mathematically, it is referred to as the $q$-Bernstein basis function, in connection with $q$-Bernstein polynomials \cite{Phillips_1997,Phillips_2010}. This $q$-deformed binomial distribution  arises, e.g., when considering the $q$-deformed generalizations of the ordinary harmonic oscillator algebra \cite{Lorek_1997}. It can be also constructed as  the probability distribution of certain events in a sequence of Bernoulli trials \cite{Charalambides_2010}.

The  probability distribution in Eq.~(\ref{q2Snu}), for $P_{2S\rar \nu}$, can also be considered as a kind of the
$q$-deformed binomial distribution. In a more general sense, a $q$-deformed binomial distribution does not have to be of the form (\ref{bbk}). It should only belong to a family of distributions that are parametrized by some parameter such that the binomial distribution is recovered at the unit value to this parameter.
Physically, in our case, the limit of the deformation parameter $x\rightarrow 1$ becomes nontrivial in the case of a large number of bosons in the
system ($N_B \gg 2S$), such that $e^{-2\pi g^2N_B} \sim O(1)$.
In this case, we can disregard  variation of bosons in the system and safely assume that all spins have independent dynamics. Then, independently of initial conditions, the distribution of the number of spin flips after the sweep of the frequency is binomial. Since the distribution in Eq.~(\ref{q2Snu}) transforms into binomial at $x\rightarrow 1$, it qualifies to be called
$q$-deformed binomial, despite it is different from (\ref{bbk}).


\section{Limits of the exact solution}
When $P_{0\rar \nu}$ and $P_{2S\rar \nu}$ are written  in terms of $q$-Pochhammer symbols and $q$-binomial coefficients, their properties still look obscure.  To simplify their appearance further, we will consider the limit $2S\gg1$. It is expected then that
probabilities of transitions to states with nearby indexes are close in magnitude, which justifies continuous approximation that replaces discrete index $\nu$ with a continuous variable. In this section, for simplicity, we will focus on the  case with $N_B=0$.




\subsection{Continuous limit}


Let us first consider the forward process. At $N_{B}=0$, Eq.~\eqref{q0nu} reduces to:
\begin{align}
&P_{0\rar \nu}=x^{2S-\nu}(x^{2S-\nu},x)_{\nu}.
\end{align}
It is convenient to introduce a new index $f\equiv2S-\nu$, 
and define
\begin{align}\label{Pf}
\mathcal{P}_f\equiv P_{0\rar (2S-f)}=x^{f}(x^{f},x)_{2S-f}.
\end{align}
Using the definition of Pochhammer symbol, one can find that   $\mathcal{P}_{f}$ satisfies the following recursion relation:
\begin{align}\label{recurPf}
&\mathcal{P}_{f+1}=\frac{x}{1-x^{f+1}}\mathcal{P}_{f},
\end{align}
which we can rewrite as
\begin{align}\label{recurPf2}
&2\frac{\mathcal{P}_{f+1}-\mathcal{P}_{f}}{\mathcal{P}_{f+1}+\mathcal{P}_{f}}
=2\left(\frac{2x}{x+1-x^{f+1}}-1\right).
\end{align}
Treating $f$ as a continuous variable, the left hand side of Eq.~\eqref{recurPf2} can be replaced by $\frac{1}{\mathcal{P}}\frac{d\mathcal{P}}{df}$, which leads to  the differential equation
\begin{align}\label{diffPf}
&\frac{1}{\mathcal{P}}\frac{d\mathcal{P}}{df}|_{f+\frac{1}{2}}=2\left(\frac{2x}{x+1-x^{f+1}}-1\right),
\end{align}
with a solution
\begin{align}\label{solPf2}
&\mathcal{P}(f)=Ce^{-\frac{(2f-1)(1-x)}{1+x}}\left(1+x-x^{f+\frac{1}{2}}\right)^{-\frac{4x}{(1+x)\log x}},
\end{align}
where $C$ is a coefficient to be determined by normalization. Interestingly, apart from $C$, $\mathcal{P}_f$ does not depend on $S$.

\begin{figure}[!htb]
\scalebox{0.33}[0.33]{\includegraphics{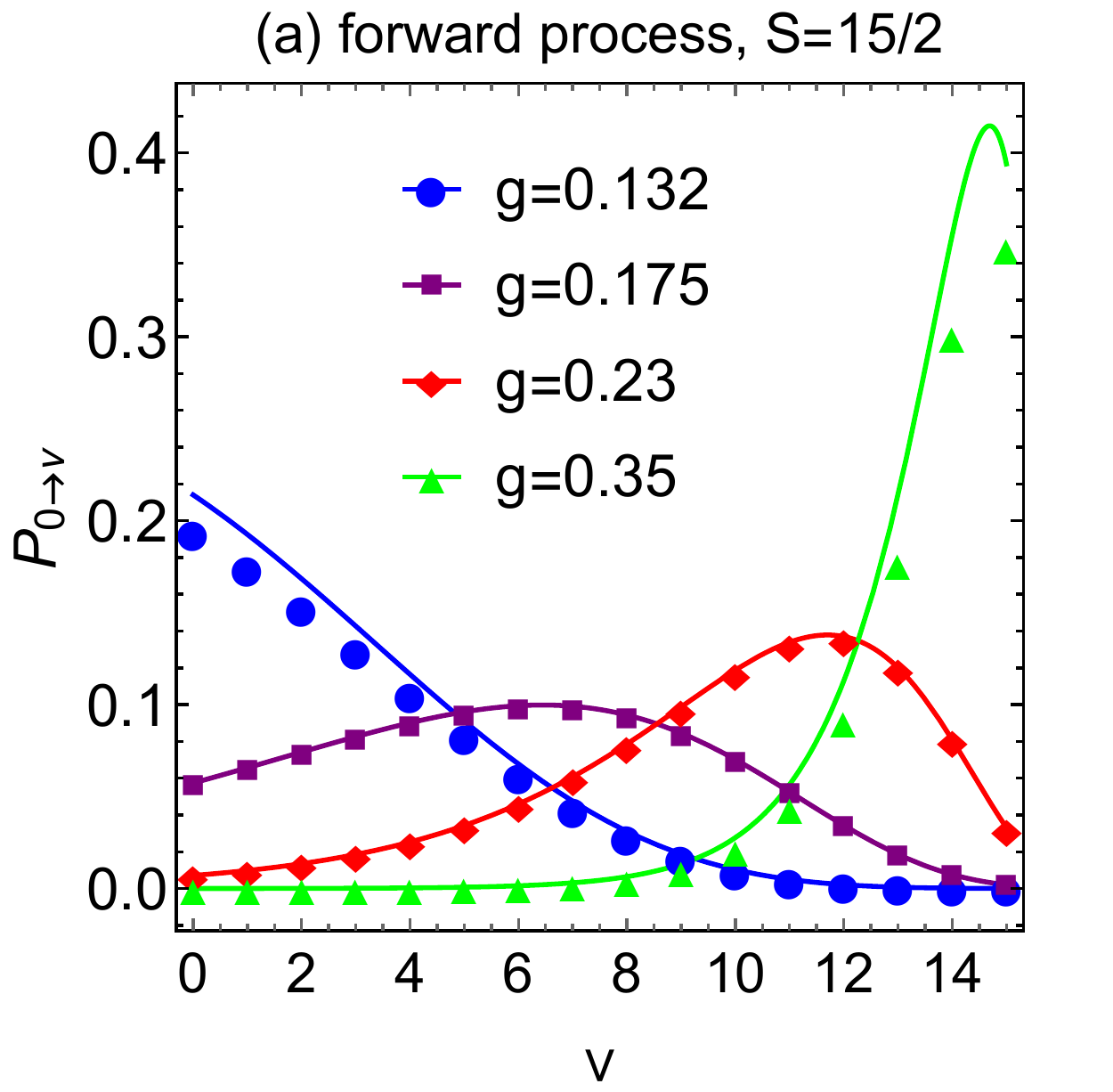}}
\scalebox{0.34}[0.34]{\includegraphics{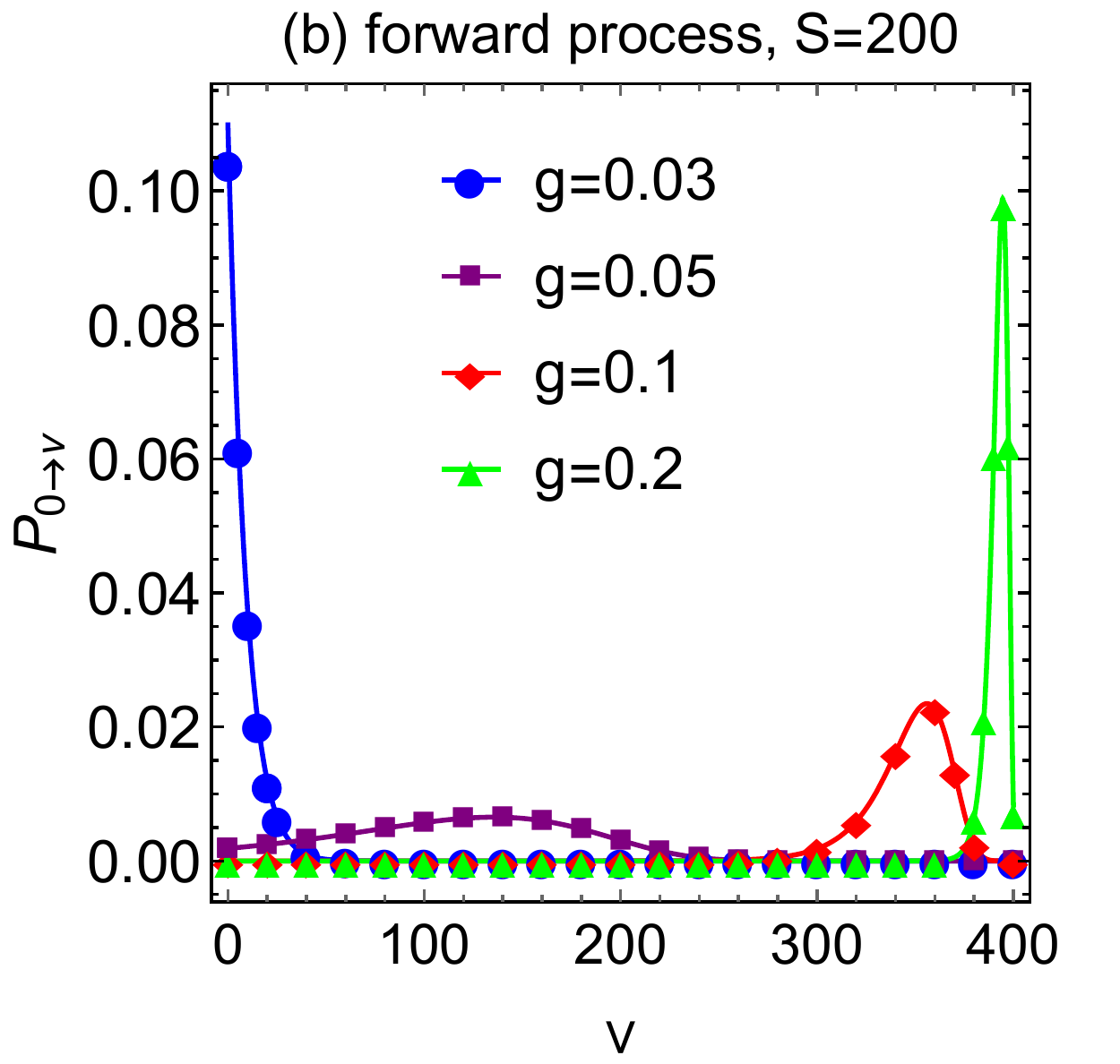}}
\scalebox{0.33}[0.33]{\includegraphics{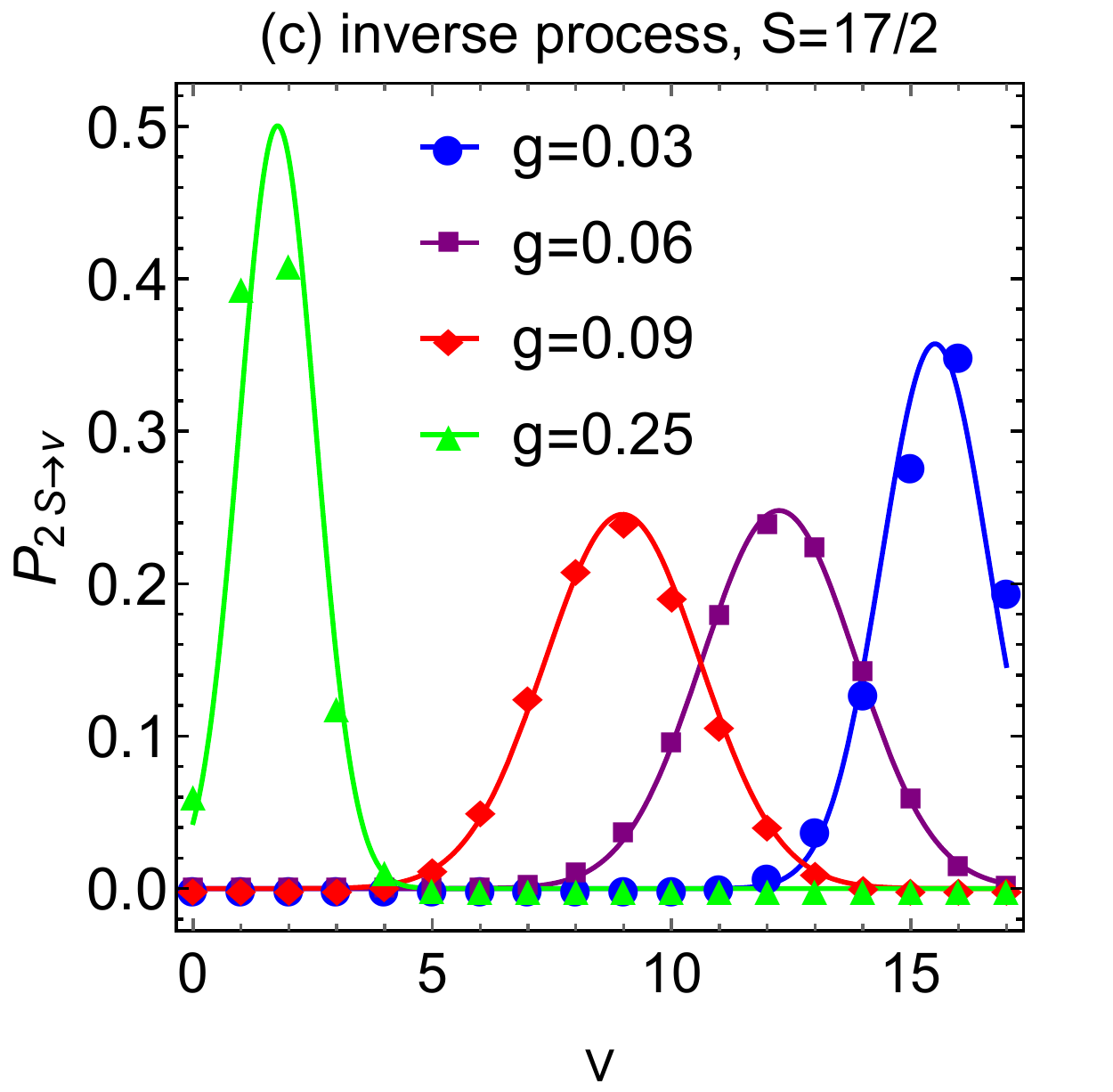}}
\scalebox{0.34}[0.34]{\includegraphics{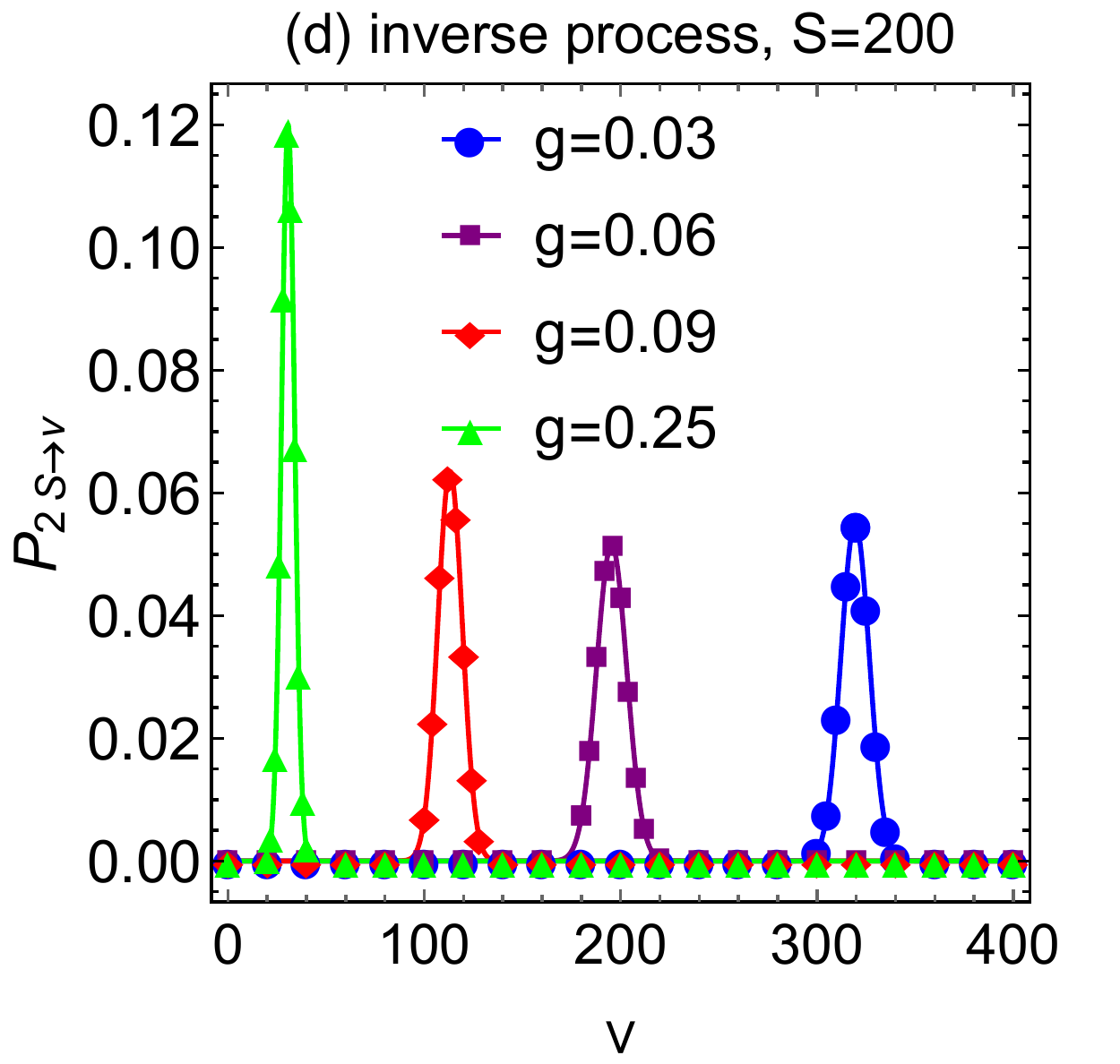}}
\caption{(color online) Comparison of exact predictions (discrete dots) and continuous approximations (solid curves) for transition probabilities in (a-b) the forward and (c-d) the inverse processes at different values of $S$'s and $g$'s, and at $N_B=0$. (a-b) For the forward process, the exact prediction is  Eq.~\eqref{q0nu}, and Eq.~\eqref{solPnu} is its continuous approximation. As $g$ increases, the curves shift toward larger $\nu$ values.
 (c-d) For the inverse process, the exact prediction is Eq.~\eqref{q2Snu}, and  its continuous limit is well represented by the Gaussian approximation in Eq.~\eqref{P2Snor}.  As $g$ increases, the curves shift to smaller $\nu$ values.
}
\label{fig1}
\end{figure}

Switching variables from $f$ back  to $\nu$, we get a continuous approximation for $P_{0\rar \nu}$:
\begin{align}\label{solPnu}
&P_{0\rar\nu}
\approx Ce^{\frac{2(1-x)}{1+x}\nu}(1+x-x^{2S+\frac{1}{2}-\nu})^{-\frac{4x}{(1+x)\log x}}.
\end{align}
Figure~\ref{fig1}(a-b) shows that   continuous approximation \eqref{solPnu} provides a very good fit to exact results at large $S$. Even for relatively small value $S=15/2$, the fit is still reasonably good (Fig.~\ref{fig1}(a)), with main deviations happening when the maximum of the distribution is at the boundary $\nu=0$ or $\nu=2S$.


From the continuous approximation \eqref{solPnu}, we can further construct a Gaussian approximation by  writing Eq.~\eqref{solPnu} as a single exponent, then finding the maximum value of the argument of this exponent, and then deriving quadratic approximation around this maximum. For the forward process the result is
\begin{align}\label{Pfnor}
&P_{0\rar\nu}\approx \frac{1}{\delta_0\sqrt{2\pi }}e^{-\frac{(\nu-\nu_0^*)^2}{2\delta_{0}^2}},
\end{align}
where the mean $\nu_{0}^*$ and variance $\delta_{0}^2$ are given by
\begin{align}\label{}
&\nu_0^*=2S-\frac{\log{(1-x)}}{\log x},\label{averforw}\\
&\delta_0^2=-\frac{ x}{(1-x)\log x}.
\end{align}
We found that this Gaussian approximation has limited applicability because the distribution $P_{0\rar \nu}$ is broad and asymmetric. So, for practical purposes, we recommend to work with a more precise approximation (\ref{solPnu}).

Analogous Gaussian approximation for the inverse process, however, turns out to be very precise. At $N_B=0$, we have
\begin{align}\label{}
 &P_{2S\rar \nu}
=\left[\begin{array}{c}
2S\\\nu
\end{array}\right]_xx^{\nu^2}(x^{\nu+1};x)_{2S-\nu},\label{}
\end{align}
which leads to the recursion relation
\begin{align}\label{recurinv}
&P_{2S\rar \nu+1}=\frac{x^{2\nu+1}-x^{2S+\nu+1}}{(1-x^{\nu+1})^2}P_{2S\rar \nu}.
\end{align}
Letting $\nu$ be a continuous variable, the differential equation for the function $P_{2S}(\nu)$, i.e., for the continuous approximation of $P_{2S\rar \nu}$, reads:
\begin{align}\label{diff1}
&\frac{1}{P_{2S}}\frac{dP_{2S}}{d\nu}\vert_{\nu+\frac{1}{2}}=2\left[\frac{2(x^{2\nu+1}-x^{2S+\nu+1})}{x^{2\nu+1}-x^{2S+\nu+1}+(1-x^{\nu+1})^2}-1 \right].
\end{align}

Explicit solution of Eq.~(\ref{diff1}) is shown in appendix. Despite its complexity, its Gaussian approximation has a simple form:
\begin{align}\label{P2Snor}
&P_{2S\rar\nu}\approx \frac{1}{\delta_{2S}\sqrt{2\pi}}e^{-\frac{(\nu-\nu_{2S}^*)^2}{2\delta_{2S}^2}},
\end{align}
with the mean $\nu_{2S}^*$ and the variance $\delta_{2S}^2$ given by 
\begin{align}\label{}
\nu_{2S}^*=-\frac{\log(2-x^{2S})}{\log x},\label{aver2S}\\
\delta_{2S}^2=-\frac{(1- x^{2S})^2}{(2-x^{2S})^2\log x}.\label{var2S}
\end{align}
Comparing with exact results shown in  Fig.~\ref{fig1}(c,d), we find that  approximation \eqref{P2Snor} works quite well and not only when $S$ is large but also for moderate spin sizes ($S\sim 10$).

\subsection{Large coupling limit}
The extremely large $g$ case  (when $x\ll1$) needs special treatment because continuous approximation cannot be justified. In this case, the distributions are dominated by a relatively small number of states near
the boundaries.
We will use condition $x\ll1$ to our advantage in order to obtain the large coupling limit of the exact result.

For the forward process at large $g$,  distribution $P_{0\rar\nu}$ is concentrated near the boundary at $\nu=2S$, so it is better to switch to the index $f\equiv 2S-\nu$ that describes the distance from this boundary.
Rewriting Eq.~\eqref{Pf} as
\begin{align}\label{Pf2}
\mathcal{P}_f\equiv P_{0\rar (2S-f)}=x^{f}\frac{(x,x)_{2S}}{(x,x)_{f-1}},
\end{align}
we recognize this as almost the Euler distribution \cite{Kemp_1992}. The only difference is due to the cut at $f=2S$, while Euler distribution extends to arbitrary nonnegative integer values of $f$.
However, in the large-$g$ limit, $\mathcal{P}_{f}$ is negligible at index values near $f=2S$, so up to exponentially small corrections, the limit of exact prediction becomes just the
 Euler distribution:
\begin{align}\label{PfEuler}
\mathcal{P}_f \approx x^{f}\frac{(x,x)_{\infty}}{(x,x)_{f-1}}.
\end{align}
The average of the Euler distribution is known \cite{Kemp_1992}:
\begin{align}\label{faverEuler}
\la f\ra=\sum_{i=1}^{\infty}\frac{1}{x^{-j}-1}=\frac{\log(1-x)+\psi_{1/x}(1)}{\log x}-\frac{1}{2},
\end{align}
where $\psi_{q}(z)$ is the $q$-digamma function defined as:
\begin{align}\label{digamma}
\psi _q(z)=\log (q) \sum _{n=0}^{\infty } \frac{q^{n+z}}{1-q^{n+z}}-\log (1-q),
\end{align}
from which we can obtain the average number of bosons generated by the forward process:
\begin{align}\label{averEuler}
n_b=2S-\la f\ra=2S+\frac{1}{2}-\frac{\log(1-x)+\psi_{1/x}(1)}{\log x}.
\end{align}


As for the inverse process,  the formula \eqref{P2Snor} works well even in the domain of large $g$, so improvements are  relevant only in the extreme limit when the average number of surviving bosons is $n_b \sim 1$. In the latter case, we can treat $2S$ and $2S-\nu$ to be $\infty$ in Eq.~\eqref{q2Snu} and get the large-$g$ limit (at $N_B=0$) as
\begin{align}\label{inverselargeg}
&P_{2S\rar \nu}
\approx x^{\nu^2}\frac{(x;x)_\infty }{[(x;x)_{\nu}]^2}.
\end{align}

\subsection{Comparison with prior theoretical estimates}

\begin{figure}[!htb]
\scalebox{0.535}[0.535]{\includegraphics{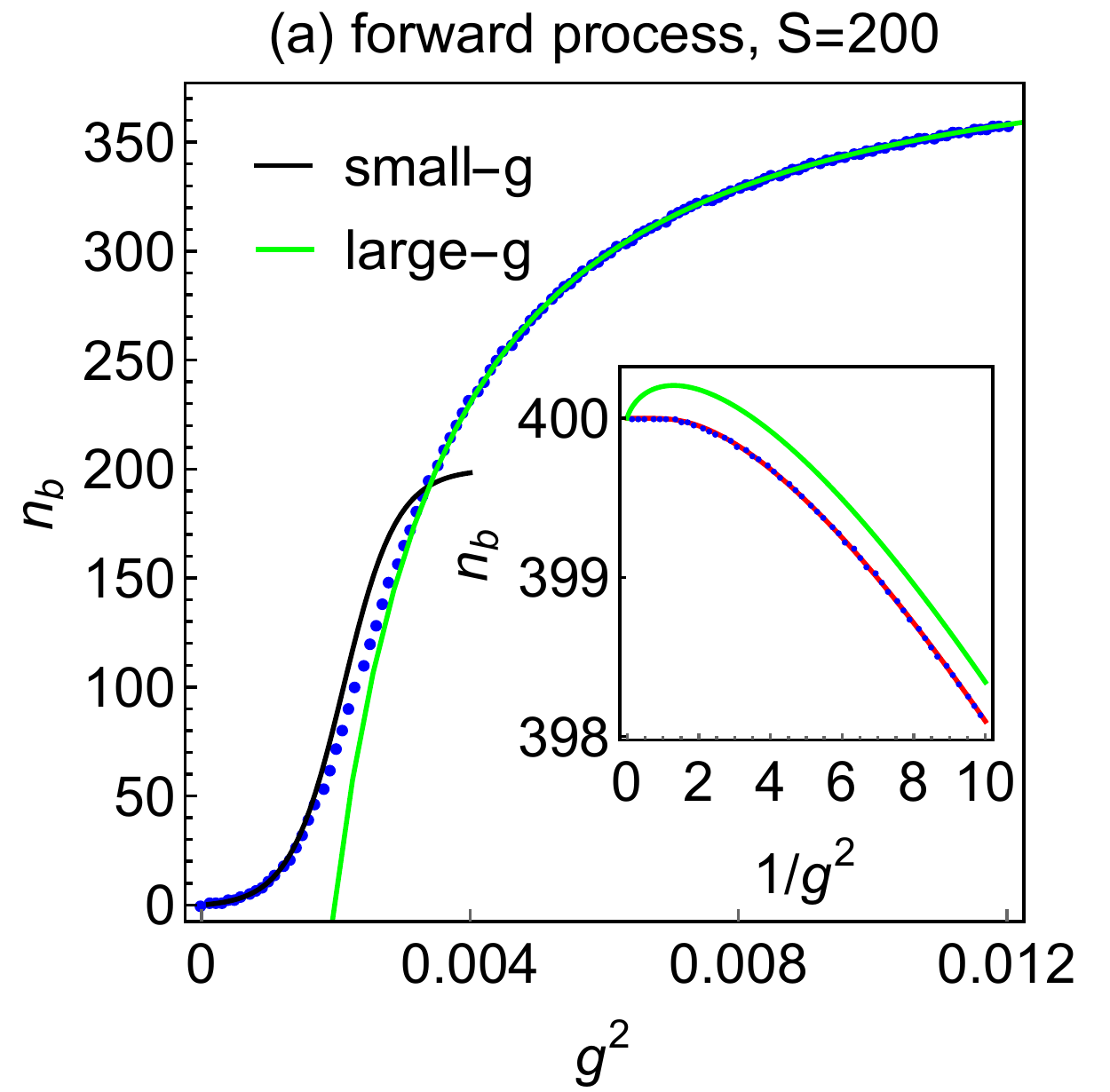}}
\scalebox{0.535}[0.535]{\includegraphics{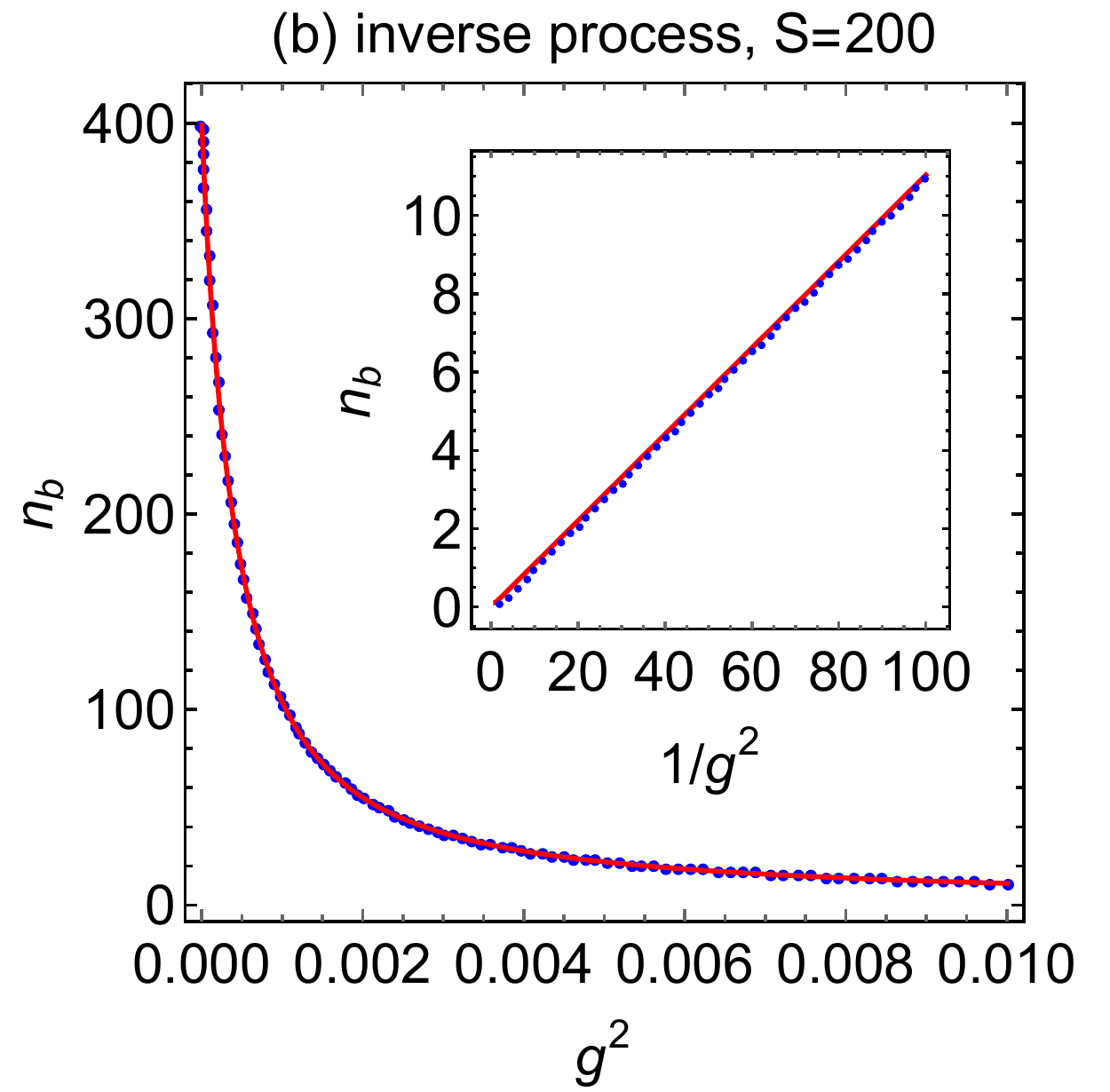}}
\caption{(color online) Average final number of bosons $n_b$ vs. square of the coupling, $g^2$, at $S=200$ and $N_B=0$ for (a)  forward and (b) inverse processes. Exact predictions of Eqs.~(\ref{q0nu})-(\ref{q2Snu})  are shown as discrete dots, and approximations  are marked as solid curves. (a) For the forward process, the black curve shows prediction \eqref{41} (Eq.~(32) in Ref. \cite{Altland_2009}) that captures the behavior  at small and moderate $g$ values ($g<1/(2S)$). The green (light gray in the grayscale version) curve shows prediction \eqref{Itininvlargeg} (Eq.~(15) in Ref. \cite{Itin_2010}) that was derived to describe the large-$g$ limit ($g>1/(2S)$). Our own approximation in Eq.~\eqref{averEuler} is shown only in the inset (red) because it would almost coincide with the green curve in the main part of the same figure.  The inset shows $n_b$ vs. $1/g^2$ for the extremely large $g$ region. It demonstrates that  Eq.~\eqref{Itininvlargeg} (green) deviates from the exact result, although this difference is suppressed by a factor $\sim 1/S$.  In contrast, our approximation \eqref{averEuler} shows no visible deviations from the exact prediction in this case at all. (b) For the inverse process, our prediction in Eq.~\eqref{averinv1} (red) fits the exact results very well at almost all values of $g$. The inset shows that $n_b$ is growing linearly with $1/g^2$ at large $g$ values ($g>1/(2S)$). Our result \eqref{averinv1} produces quantitatively the same prediction for this slope as the prior semiclassical prediction of Ref.~\cite{Itin_2009}.}
\label{fig2}
\end{figure}
With our formulas \eqref{q0nu} and \eqref{q2Snu}, we can check validity of the prior theoretical predictions  \cite{Altland_2008,Altland_2009,Itin_2009,Itin_2010} that were derived to describe  the average number of bosons $n_b=\langle \nu \rangle$ after the frequency chirp.
For the forward process, Ref.~\cite{Altland_2009} used diagrammatic kinetic approach to derive a formula (Eq.~(31) in \cite{Altland_2009}) for the average number of created bosons at small and moderate couplings ($g<1/(2S)$):
\begin{align}
&n_b\approx \frac{2S(x^{-2S}-1)}{x^{-2S}+2S}, \qquad \textrm{for }g<\frac{1}{2S},
\label{41}
\end{align}
In Ref.~\cite{Itin_2010}, the opposite, i.e. the large-$g$ limit of the forward process was considered by mapping the semiclassical nonlinear evolution equation to the Painlev\'e-II equation with known asymptotics. As a result, the following formula (rewritten in our notation) was proposed:
\begin{align}
&n_b=2S-\frac{1}{\log x}[\log (-\log x)-\gamma], \qquad \textrm{for }g>\frac{1}{2S},
\label{Itininvlargeg}
\end{align}
where $\gamma$ is the Euler's constant.

In Fig.~\ref{fig2}(a) we plot predictions of Eqs.~(\ref{41})-(\ref{Itininvlargeg}) showing that they, indeed, work quite well within their domains of validity. Our own prediction for the large-$g$ limit ($g>1/(2S)$), given in Eq.~\eqref{averEuler}, also works  well but we show it only in the inset to Fig.~\ref{fig2}(a) because our prediction almost coincides with prediction of Eq.~(\ref{Itininvlargeg}). The difference is only in terms that are suppressed by a factor $\sim 1/S$, which is expected because our approximation includes quantum effects at early stages of the dynamics, which cannot be done within the semiclassical approach.

The inverse process was studied  in \cite{Altland_2009} and \cite{Itin_2009} in the large-$g$ limit ($g>1/(2S)$). These articles predicted that $n_b$ should depend linearly on $1/g^2$ in this limit, with some difference in predictions for the proportionality coefficient. 
 We can also derive this linear dependence of $n_b$ on $1/g^2$.
Our formula \eqref{aver2S} makes the following prediction
\begin{align}\label{averinv1}
n_b=-\frac{\log(2-x^{2S})}{\log x},
\end{align}
which is valid up to the values of $g \sim 1$, i.e., when the number of remaining bosons is $n_b\sim1$. For small values of $n_b$ it predicts
\begin{align}\label{invlinear}
n_b\approx-\frac{\log2}{\log x} =\frac{\log2}{ 2\pi g^2}.
\end{align}
The obtained value of the coefficient, $(\log2)/\pi$, coincides with its semiclassical prediction in Ref.~\cite{Itin_2009}. The additional factor $1/2$ in our result is  due to the difference in notation.
In Fig.~\ref{fig2}(b) we verify validity of both formulas (\ref{averinv1})-(\ref{invlinear}).
Thus we come to conclusion that 
semiclassical approximations that were developed in the series of publications \cite{Altland_2008,Altland_2009,Itin_2009,Itin_2010} capture both qualitative and quantitative behavior of the mean value of the boson number distribution. 

\section{Conclusion}

We simplified the exact solution of the  Tavis-Cummings model with a linearly time-dependent bosonic mode frequency.
 For the nondegenerate model (\ref{HamiltonianTC}), we found a general formula for the monomial that describes the transition probability between two arbitrary diabatic states. This formula explains all observations about the exact solution that were made in \cite{Sinitsyn_2016}. We showed then that, for fully spin polarized initial conditions, probabilities to emit a given number of bosons can be expressed  in this model in terms of $q$-deformed binomial distributions, which admit  simple continuous approximations in the case of many spins.

Comparisons of  different limits of exact results to previous predictions that were based on advanced semiclassical and diagrammatic techniques \cite{Altland_2008,Altland_2009,Itin_2009,Itin_2010} confirm the validity of methods developed in those publications.
Especially astounding  is the agreement with semiclassical predictions    that were obtained  in \cite{Itin_2009,Itin_2010} for large couplings by mapping dynamics  to the Painlev\'e-II equation. There is practically no visible deviation of those predictions from exact results within the large-$g$ range, as shown in Fig.~\ref{fig2}.  Moreover,  the quantitative estimate of the proportionality coefficient of the slope in the inset of Fig.~\ref{fig2}, which was derived in \cite{Itin_2009}, turns out to be the same as the one that we derived from the exact solution of the model. Apparently, this agreement with semiclassical calculations means that purely quantum effects that are essential at early stages of the process have only minor influence on final state probabilities in the strong coupling limit.  Availability of the exact solution, however, allowed us to extend results in Refs.~\cite{Altland_2008,Altland_2009,Itin_2009,Itin_2010} and  explore small purely quantum effects rigorously, describe the intermediate coupling regime, and obtain other than mean characteristics of distributions.

The standard notion of integrability in quantum mechanics does not normally cover the scope  of problems with explicit time-dependence of parameters and a combinatorially large size of the phase space. Solution of the driven Tavis-Cummings model in \cite{Sinitsyn_2016} is the proof that extensions of the notion of quantum integrability to this new domain of problems is possible.

Our present article suggests that  quantum algebras may provide the basic framework for such an extension because $q$-deformed distributions that we found arise  typically in relation to such algebras.
Other indications include the fact that the time-independent version of the Tavis-Cummings model is known to be solvable by the algebraic Bethe ansatz \cite{tv-bethe}. Moreover, the relation to Painlev\'e-II equation, which was demonstrated in \cite{Itin_2009,Itin_2010} and supported by our work, suggests another possible way to understand integrability in the multistate LZ theory. For example, linearizations of Painlev\'e equations correspond to nonstationary Schr\"odinger equations with time-dependent coefficients; for Painlev\'e-I and Painlev\'e-II such a linearization has the form similar to (\ref{mlz}) \cite{zabrodin}. All such observations suggest that multistate LZ integrability has the same mathematical roots as the conventional quantum and classical integrability. We hope that our work will stimulate interest in understanding this relation.

\section*{Acknowledgements}
The work
was carried out under the auspices of the National Nuclear
Security Administration of the U.S. Department of Energy at Los
Alamos National Laboratory under Contract No. DE-AC52-06NA25396. Authors also thanks the support from the LDRD program at LANL.

\section*{Appendix}

\renewcommand{\theequation}{A1}

Explicit solution of Eq.~(\ref{diff1}) is given by
\begin{align}\label{}
&P_{2S\rar \nu}(\nu)=&\\
\nonumber &Ce^{-2 \nu +\frac{4 \left(x^{2 S+\frac{1}{2}}+2 x^{2 S+\frac{3}{2}}-2 \sqrt{x}\right) \arctan\left(\frac{x^{2 S+\frac{1}{2}}-2 x^{\nu +1}-2 x^{\nu }+2 \sqrt{x}}{\sqrt{-4 x^{2 S+1}-x^{4 S+1}+4}}\right)}{(x+1) \sqrt{-4 x^{2 S+1}-x^{4 S+1}+4} \log (x)}}& \\
\nonumber &\times \left(1-x^{\nu +2 S+\frac{1}{2}}+x^{2 \nu }-2 x^{\nu +\frac{1}{2}}+x^{2 \nu +1}\right)^{\frac{2}{(1+x) \log (x)}}&,
\end{align}
where $C$ is a normalization coefficient.

\end{document}